 \documentclass[aps,twocolumn,pra,superscriptaddress,amsmath,showpacs,tightenlines]{revtex4}

\usepackage{epsfig,graphicx,times}
\usepackage{amstext}
\usepackage{amsmath}            
\usepackage{amssymb}            
\usepackage{graphicx}           
\usepackage{latexsym}
\usepackage{bm}

\newtheorem{criterion}{Criterion}
\newtheorem{definition}{Definition}

\newcommand{\ncl}{{\;\rm\raisebox{-2.pt}{\,$\stackrel{\rm ncl}{<}\;$}}}
\newcommand{\nclg}{{\;\rm\raisebox{-2.pt}{\,$\stackrel{\rm ncl}{>}\;$}}}
\newcommand{\ent}{{\;\rm\raisebox{-2.pt}{\,$\stackrel{\rm ent}{>}\;$}}}
\newcommand{\cl}{{\;\rm\raisebox{-2.pt}{$\stackrel{\rm cl}{\ge}\;$}}}

\newcommand{\jpa}{J. Phys. A~}
\newcommand{\jpb}{J. Phys. B~}
\newcommand{\pla}{Phys. Lett. A~}
\newcommand{\job}{J. Opt. B: Quantum Semiclass. Opt.~}

\def\pt#1{{\langle  #1 \rangle^{\Gamma}}}
\def\normal#1{{\langle :  #1 : \rangle}}

\newcommand\Mat[4]{{\left(\begin{array}{cccc} #1 \\ #2  \\ #3  \\ #4 \end{array} \right)}}

\begin{document}

\title{Sudden vanishing and reappearance of nonclassical effects:
General occurrence of finite-time decays and periodic
vanishings of nonclassicality and entanglement witnesses}

\author{Monika Bartkowiak} 
\affiliation{Faculty of Physics, Adam Mickiewicz University,
PL-61-614 Pozna\'n, Poland}

\author{Adam Miranowicz}
\affiliation{Faculty of Physics, Adam Mickiewicz University,
PL-61-614 Pozna\'n, Poland} \affiliation{Advanced Science
Institute, RIKEN, Wako-shi, Saitama 351-0198, Japan}

\author{Xiaoguang Wang}
\affiliation{Advanced Science Institute, RIKEN, Wako-shi, Saitama
351-0198, Japan} \affiliation{Zhejiang Institute of Modern
Physics, Department of Physics, Zhejiang University, Hangzhou
310027, China}

\author{Yu-xi Liu}
\affiliation{Institute of Microelectronics, Tsinghua University,
Beijing 100084, China} \affiliation{Tsinghua National Laboratory
for Information Science and Technology (TNList), Tsinghua
University, Beijing 100084, China} \affiliation{Advanced Science
Institute, RIKEN, Wako-shi, Saitama 351-0198, Japan}

\author{Wies\l{}aw Leo\'nski}
\affiliation{Institute of Physics, University of Zielona G\'ora,
PL-65-516 Zielona G\'ora, Poland}

\author{Franco Nori}
\affiliation{Advanced Science Institute, RIKEN, Wako-shi, Saitama
351-0198, Japan} \affiliation{Physics Department, The University
of Michigan, Ann Arbor, Michigan 48109-1040, USA}

\date{\today}

\begin{abstract}
Analyses of phenomena exhibiting finite-time decay of quantum
entanglement have recently attracted considerable attention.
Such decay is often referred to as sudden vanishing (or
sudden death) of entanglement, which can be followed by its
sudden reappearance (or sudden rebirth). We analyze various
finite-time decays (for dissipative systems) and analogous
periodic vanishings (for unitary systems) of nonclassical
correlations as described by violations of classical
inequalities and the corresponding nonclassicality witnesses
(or quantumness witnesses), which are not necessarily
entanglement witnesses. We show that these sudden vanishings
are universal phenomena and can be observed: (i) not only for
two- or multi-mode but also for single-mode nonclassical
fields, (ii) not solely for dissipative systems, and (iii) at
evolution times which are usually different from those of
sudden vanishings and reappearances of quantum entanglement.
\end{abstract}

\pacs{42.50.Xa,03.67.Mn}

\maketitle

\pagenumbering{arabic}

\section{Introduction}

Decoherence is a crucial obstacle in practical
implementations of quantum information processing and quantum
state engineering. Quantum entanglement is especially fragile
to decoherence. Yu and Eberly~\cite{Yu04} (see also earlier
studies in Refs.~\cite{Zyczkowski}) observed that
entanglement decay can occur within a finite time. This
effect has been referred to as entanglement {}``sudden
death'' or entanglement sudden vanishing (SV) and it can be
followed by its sudden reappearance (sudden
rebirth---SR)~\cite{Zyczkowski,Tanas04,Lopez08}.
Reference~\cite{Yu04} has triggered extensive theoretical
research on entanglement loss in various systems (for reviews
see Ref.~\cite{YuEberlyReview}). Entanglement sudden
vanishing was also experimentally
observed~\cite{Almeida07,Laurat07,Xu10}.

Entanglement SV is often considered to be a new form of decay
of quantum entanglement, which presumably was not previously
encountered in the dissipation of other physical
correlations. Here we would like to point out the
\emph{general occurrence of sudden finite-time decays and
periodic vanishings of nonclassical correlations}. Namely,
the SV and SR effects can also be observed during the
evolution of entanglement
witnesses~\cite{Horodecki96,Hyllus,Guhne} (for a review see
Ref.~\cite{Horodecki-review}) and nonclassicality witnesses
(also called quantumness
witnesses)~\cite{SRV,Korbicz,Alicki,Zukowski,Semenov,Filippov,Miran10}
corresponding to violations of classical inequalities.

A standard approach to study the SV and SR of quantum
entanglement is based on the analysis of the time evolution
of entanglement measures, e.g., the concurrence or,
equivalently, the negativity or the relative entropy of
entanglement~\cite{Horodecki-review}. For a two-qubit system,
described by a density matrix $\hat{\rho}$, the concurrence
$C(\hat{\rho})$ is defined by~\cite{Wootters}:
\begin{equation}
C(\hat{\rho})=\max\Big(0,2\max_{i}\lambda_{i}-\sum_{i}
\lambda_{i}\Big),\label{concurrence}
\end{equation}
where the $\lambda_{i}$'s are the square roots of the eigenvalues
of $\hat{\rho}(\hat{\sigma}_{2}\otimes\hat{\sigma}_{2})
\hat{\rho}^{*}(\hat{\sigma}_{2}\otimes\hat{\sigma}_{2})$ and
$\hat{\sigma}_{2}$ is the Pauli spin matrix. On the other hand,
the negativity can be defined as~\cite{Peres,Horodecki96}:
\begin{equation}
N(\hat\rho)=\max\Big(0,-2\min_j \mu _{j}\Big), \label{negativity}
\end{equation}
where $\mu _{j}$'s are the eigenvalues of the partial transpose
$\hat\rho^{\Gamma}$ and factor 2 is chosen for proper scaling,
i.e., to get $N(\hat\rho)=1$ for Bell's states.

It is worth noting that not all the SRs and SVs of
entanglement and its witnesses can be considered standard: A
SR should appear only after some finite-evolution time after
the occurrence of the preceding SV. Specifically, let us now
analyze an example: Both $|\cos t|$ and $\max(0, \cos t)$
vanish at $\pi/2$, but only the vanishing of the latter
function is associated with the proper SV and SR effects.

Both Eqs.~(\ref{concurrence}) and~(\ref{negativity}) are given as
the maximum of zero and some functions, which clearly explains the
occurrence of SVs if $\hat{\rho}$ changes in time. By contrast,
SVs do not appear for the modified parameters
$C'(\hat{\rho})=2\max_{i}\lambda_{i}-\sum_{i}\lambda_{i}$, and
$N'(\hat{\rho})=-2\min_j \mu _{j}$, if $\lambda_{i}$ and $\mu
_{j}$ have continuous derivatives in time.

We deduce that analogous SV and SR effects can be observed for an
arbitrary time-dependent parameter $F(t)$, in comparison to some
threshold value $F_{0}$. From a quantum-mechanical point of view,
the most interesting parameters $F$ are the ones which correspond
to classical inequalities $F\cl F_{0}$ that \emph{can} be violated
for some \emph{nonclassical} fields, i.e., $F\ncl F_{0}$ as
indicated by the symbol $\ncl$. On the other hand, the symbol
$\cl$ emphasizes that the corresponding inequality \emph{must} be
fulfilled for all \emph{classical} states. Thus, let us truncate
such parameter $F$ as follows:
\begin{eqnarray}
F\rightarrow\tilde{F}=\max(0,F_{0}-F).\label{redefine}
\end{eqnarray}
A simple illustration of this concept is shown in Fig.~1. For
brevity, such $F$ and $\tilde{F}$ will be referred to as the
untruncated and truncated \emph{nonclassicality witnesses},
respectively. The redefinition of the witnesses is a key
concept in observing the SV and SR effects.

\begin{figure}
\centerline{\epsfxsize=8.5cm\epsfbox{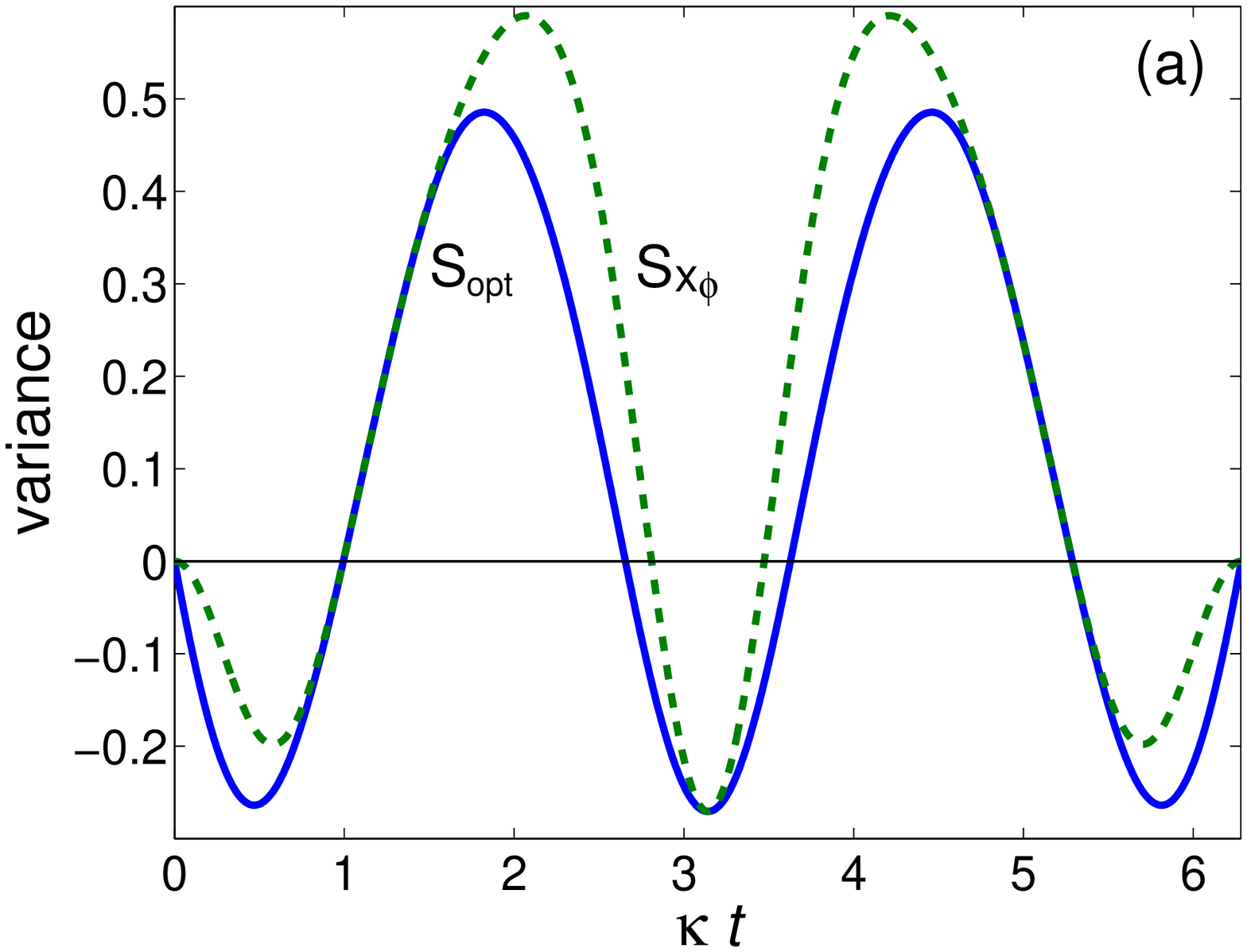}}
\centerline{\epsfxsize=8.5cm\epsfbox{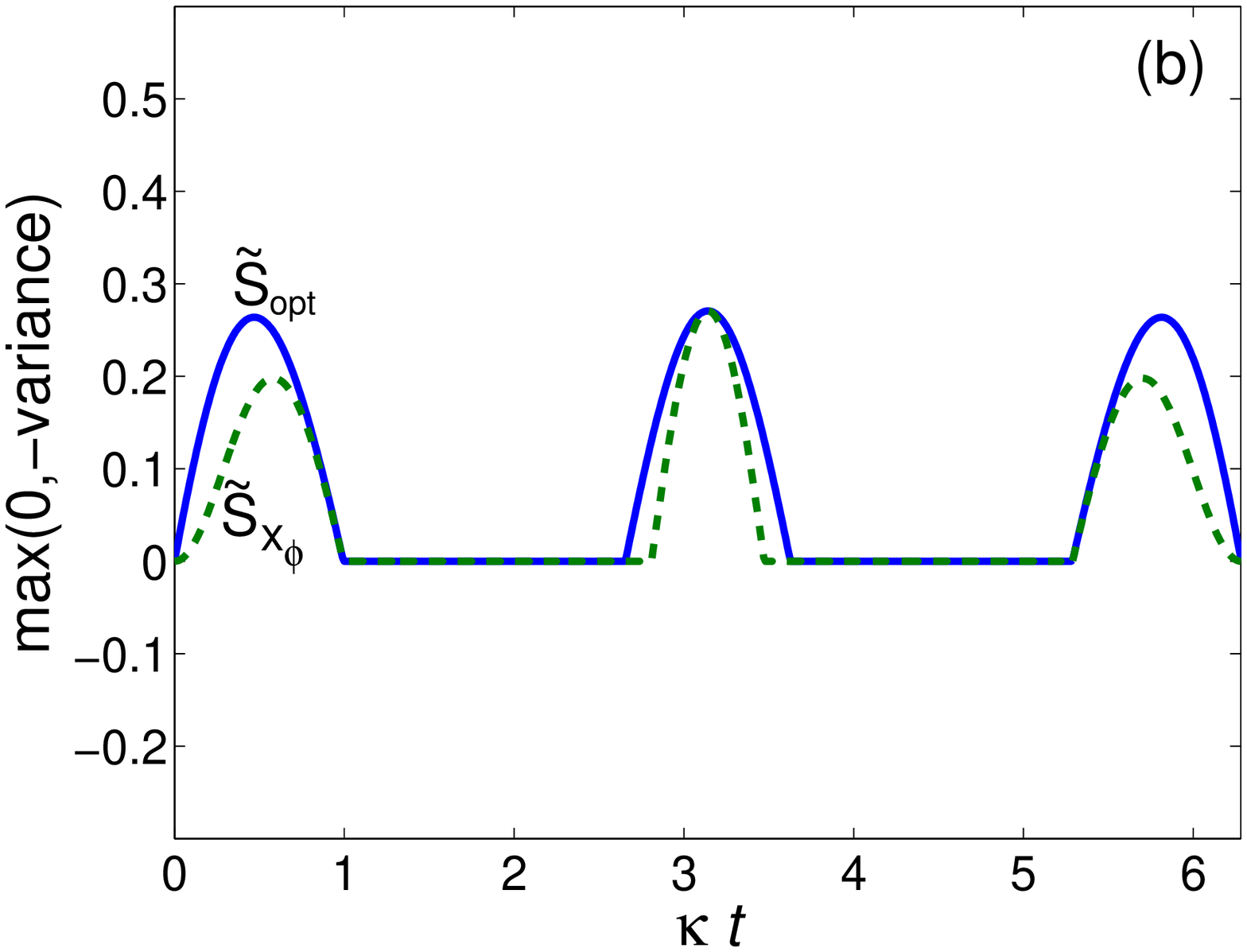}}
\caption{(Color online) A simple explanation of how to
observe the SV and SR of nonclassicality witnesses using, as
an example, the unitary evolution of single-mode squeezing in
the anharmonic oscillator model given by the
Hamiltonian~(\ref{a1}): (a) normally-ordered variances
$S_{x_{\phi}}$ (dashed curve) and $S_{\rm opt}$ (solid
curve), given by Eqs.~(\ref{SxPhiAOM}) and~(\ref{SoptAOM}),
(b) truncated normally-ordered variances $\tilde
S_{x_{\phi}}$ (dashed curve) and $\tilde S_{\rm opt}$ (solid
curve), given by Eqs.~(\ref{SxPhiBar}) and~(\ref{SoptBar}),
respectively. Quadrature squeezing occurs if $S_{x_{\phi}}<
0$ or, equivalently, if the truncated witness
$\tilde{S}_{x_{\phi}}> 0$. Principal squeezing occurs if
$S_{\rm opt}< 0$ or if the truncated witness $\tilde{S}_{\rm
opt}> 0$. Here, $|\alpha_0|^2=1/2$, $\phi_0=\phi=0$, and
$S_0=0$. By including damping, one would observe finite-time
decays, analogously to the standard sudden decays of
entanglement.}
\end{figure}

In the next sections we give general arguments and present
some specific examples of phenomena and nonclassicality
witnesses to support our conclusions.

In Sec. II we recall a definition of nonclassicality and
present a general method of constructing truncated
nonclassicality witnesses that can exhibit both the SV and SR
effects. In Sec. III, we discuss methods of constructing
truncated entanglement witnesses. We also give a few simple
examples of truncated nonclassicality and entanglement
witnesses. Their evolution in some prototype physical models
is studied in Secs. IV-VI. We conclude in Sec. VII.

\section{Nonclassicality witnesses}

In order to test (and characterize) the nonclassical behavior
of a given state $\hat{\rho}$ unambiguously, we use the
multimode Cahill-Glauber $s$-parametrized quasiprobability
distribution (QPD) functions defined for $-1\le s\le1$
by~\cite{Cahill}:
\begin{equation}
{\cal W}^{(s)}(\bm{\alpha})=\frac{1}{\pi}\,{\rm
Tr}\left(\hat{\rho}\,\prod_{k=1}^{M}\hat{T}^{(s)}(\alpha_{k})\right),
\label{N09}
\end{equation}
where
\begin{equation}
\hat{T}^{(s)}(\alpha_{k})=\frac{1}{\pi}\int\exp\left(\alpha_{k}\xi^{*}-\alpha_{k}^{*}\xi+\frac{s}{2}|\xi|^{2}\right)\hat{D}(\xi)\,{\rm
d}^{2}\xi,\label{N09a}
\end{equation}
$\hat{D}(\xi)$ is the displacement operator, $\bm{\alpha}$ is
a complex multivariable $(\alpha_1,\alpha_2,...,\alpha_M)$,
and $M$ is the number of modes. In special cases (for
$s=1,0,-1$), the QPD reduces to the standard
Glauber-Sudarshan $P$ function, Wigner $W$ function, and
Husimi $Q$ function, respectively.

A well-known criterion of nonclassicality (or quantumness) is
based on the $P$ function (see, e.g., Refs.~\cite{VogelBook}):
\begin{definition}
A state $\hat{\rho}$ is considered \emph{nonclassical} if its
Glauber-Sudarshan $P$ function is not a classical probability
density (i.e., it is nonpositive). Otherwise the state
$\hat{\rho}$ is called classical.
\end{definition}
We use this definition of nonclassicality although we are
aware of its drawbacks (see, e.g., Ref.~\cite{Wunsche}). It
is also worth noting that this definition is often extended
by a requirement of nonsingularity. That is, a classical $P$
function cannot be more singular than Dirac's $\delta$
function. But, in fact, the singularity of the $P$ function
is implied by its nonpositivity (see, e.g.,
Ref.~\cite{Miran10}).

Definition~1 can be equivalently formulated via a complete
set of nonclassicality witnesses corresponding to violations
of classical inequalities. Here we apply the method of
constructing nonclassicality witnesses proposed in
Refs.~\cite{SRV,Korbicz} and developed in
Refs.~\cite{Vogel08,Miran10}. Alternatively, one can apply an
approach used by Alicki \emph{et
al.}~\cite{Alicki,Zukowski,Filippov}.

Let us analyze an arbitrary $M$-mode operator
$\hat{f}\equiv\hat{f}(\hat{{\bf a}},\hat{{\bf a}}^{\dagger})$
as a function of the annihilation, $\hat{{\bf
a}}\equiv(\hat{a}_{1},\hat{a}_{2},...,\hat{a}_{M})$, and
creation, $\hat{{\bf a}}^{\dagger}$, operators. The $P$
function enables a direct calculation of the normally-ordered
(denoted by ::) expectation values of the Hermitian operator
$\hat{f}^{\dagger}\hat{f}$ as follows:
\begin{eqnarray}
\langle :\hat{f}^{\dagger} \hat{f}:\rangle &=& \int{\rm
d}^{2}\bm{\alpha}\;|f(\bm{\alpha},\bm{\alpha}^{*})|^{2}
P(\bm{\alpha},\bm{\alpha}^{*}). \label{ff}
\end{eqnarray}
Then one can apply another criterion of
nonclassicality~\cite{SRV,Korbicz}:
\begin{criterion}
A state $\hat{\rho}$ is classical if $\langle :\hat f^\dagger
\hat f :\rangle \ge 0$ for all functions $\hat f$.
Conversely, if $\langle : \hat f^\dagger \hat f :\rangle < 0$
for some $\hat f$ then the state $\hat{\rho}$ is
nonclassical.
\end{criterion}
These conditions can be compactly written as $\langle :\hat
f^\dagger \hat f :\rangle \cl 0$ and $\langle :\hat f^\dagger
\hat f :\rangle \ncl 0$. By analogy with definitions of
entanglement witness (see the following section), the
normally-ordered Hermitian operator $:\!\hat f^\dagger \hat f
\!:$ can be referred to as (nonlinear) nonclassicality (or
quantumness) witness~\cite{Korbicz}. For convenience, we call
the nonclassicality witness (and also entanglement witness)
not only an observable but also its expectation value. Note
that the understanding of nonclassicality witnesses is not
strictly limited to operators (see, e.g.,
Refs.~\cite{Filippov,Semenov}).

By writing $\hat f = \sum_{i}^N c_{i} \hat f_{i}$, where
$c_{i}$ are arbitrary complex numbers, one obtains
\begin{eqnarray}
\normal{\hat{f}^{\dagger}\hat{f}} & = &
\sum_{i,j}c_{i}^{*}c_{j}\normal{\hat{f_{i}}^{\dagger}\hat{f}_{j}}.
\label{w1}
\end{eqnarray}
The normally-ordered moments $\normal{\hat{f_{i}}
^{\dagger}\hat{f}_{j}}$ can be grouped into the following matrix:
\begin{eqnarray}
 M^{\rm (n)}_{\hat f}(\hat\rho)
= \Mat{ \normal{\hat f_{1}^\dagger \hat f_{1}} & \normal{\hat
f_{1}^\dagger \hat f_{2}} & \cdots & \normal{\hat f_{1}^\dagger
\hat f_{N}} } { \normal{\hat f_{2}^\dagger \hat f_{1}} &
\normal{\hat f_{2}^\dagger \hat f_{2}} & \cdots & \normal{\hat
f_{2}^\dagger \hat f_{N}} } { \vdots & \vdots & \ddots & \vdots }
{ \normal{\hat f_{N}^\dagger \hat f_{1}} & \normal{\hat
f_{N}^\dagger \hat f_{2}} & \cdots & \normal{\hat f_{N}^\dagger
\hat f_{N}} }.
  \label{w2}
\end{eqnarray}
We call (nonlinear) nonclassicality witnesses not only
\linebreak $:\!\hat f^\dagger \hat f \!:$ and $\langle
:\hat{f}^{\dagger} \hat{f}:\rangle$ but also the matrices of
normally-ordered moments $M^{\rm (n)}_{\hat f}(\hat\rho)$ and
their functions (e.g., determinants). The importance of this
approach is motivated by the following nonclassicality
criterion~\cite{SRV,Miran10}:
\begin{criterion}
A state $\hat\rho$ is nonclassical if there exists $\hat f$, such
that $\det[M^{\rm (n)}_{\hat f}(\hat\rho)]$ is negative.
\end{criterion}
Thus, if these nonclassicality witnesses are truncated
according to Eq.~(\ref{redefine}), one can predict infinitely
many different kinds of SV and SR effects. Note that a given
nonclassicality witness reveals only some specific and
limited properties of nonclassical states.

It is worth stressing that nonclassicality witnesses are
(usually) not measures of nonclassicality. A question arises
whether SV and SR effects can also be observed for some
nonclassicality measures. Below we give an example of quantum
dynamics leading to the SV and SR of nonclassicality
witnesses but \emph{not} of nonclassicality measures.

\subsection{Examples of truncated nonclassicality witnesses}

To find nontrivial examples of SV and SR of some
nonclassicality witnesses, which are not necessarily
entanglement witnesses (studied in the following section), we
analyze the squeezing (or sub-Poisson statistics) of the
photon-number difference $(\hat{n}_{1}-\hat{n}_{2})$ in two
systems. This squeezing occurs if the normally-ordered
variance
\begin{equation}
S=\langle:[\Delta(\hat{n}_{1}-\hat{n}_{2})]^{2}:\rangle\label{Sn0}
\end{equation}
is negative, where
$\Delta\hat{O}\equiv\hat{O}-\langle\hat{O}\rangle$, with
$\hat{O}=\hat{n}_{1}-\hat{n}_{2}$. It is a purely
nonclassical effect as $S\cl0$ holds for any classical
fields. Note that $S+S_{0}\cl 0$ also holds for any classical
fields, where $S_{0}\ge 0$ is a threshold value which can be
chosen to be arbitrary. Thus, one can analyze a kind of
{}``strong'' squeezing if $S+S_{0}\ncl 0$. In order to
observe the SV and SR of this strong squeezing we truncate
the squeezing parameter $S$ as follows:
\begin{eqnarray}
\tilde{S}=\max\big(0,-\langle:[\Delta(\hat{n}_{1}-\hat{n}_{2})]^{2}:
\rangle-S_{0}\big)\nclg0.\label{Sn}
\end{eqnarray}
By replacing $\Delta(\hat{n}_{1}-\hat{n}_{2})$ by
$(\hat{n}_{1}-\hat{n}_{2})$ in Eq.~(\ref{Sn0}), one can consider
another normally-ordered witness $\tilde{D}'$ resulting from the
classical inequality
\begin{equation}
D'=\langle:(c_{1}\hat{n}_{1}+c_{2}\hat{n}_{2}+c_{3}){}^{2}:\rangle
+|c_{4}|^{2}\cl0\label{Dnn}
\end{equation}
assuming real parameters $c_{k}$ ($k=1,2,3,4$). In the following,
we apply
\begin{equation}
\tilde{D}=\max(0,-\langle:(\hat{n}_{1}-\hat{n}_{2}
+D_{0})^{2}:\rangle)\nclg0,\label{Dn}
\end{equation}
which is a special case of $\tilde{D}'$ for
$(c_{1},c_{2},c_{3},c_{4})=(1,-1,D_{0},0)$.

So far we have only analyzed two-mode witnesses. Clearly, it
is also possible to observe the SV and SR during the time
evolution of multi-mode but also single-mode witnesses of
nonclassicality. We give only two examples of photon-number
and quadrature squeezings:

(i) Single-mode photon-number squeezing (also called sub-Poisson
photon-number statistics) occurs if Mandel's $Q$-parameter is
negative, i.e.,~
$\langle:(\Delta\hat{n})^{2}:\rangle/\langle:\hat{n}:\rangle\ncl0$.
This nonclassical effect can also be described by the truncated
witness
\begin{eqnarray}
\tilde{Q}=\max\left(0,-\,\frac{\langle:(\Delta\hat{n})^{2}:\rangle}
{\langle:\hat{n}:\rangle}\right)\nclg0.\label{N17}
\end{eqnarray}
(ii) The standard ($S_{0}=0$) and strong ($S_{0}>0$) $M$-mode
quadrature squeezing can be defined by
\begin{equation}
S_{x_{\phi}}=\langle:(\Delta\hat{x}_{\bm{\phi}})^{2}:\rangle
\ncl(-S_{0}),\label{SxPhi}
\end{equation}
or, equivalently, via the truncated squeezing witness
\begin{equation}
\tilde{S}_{x_{\phi}}=\max(0,
-\langle:(\Delta\hat{x}_{\bm{\phi}})^{2}:\rangle
-S_{0})\nclg0,\label{SxPhiBar}
\end{equation}
where $\bm{\phi}=(\phi_{1},\phi_{2},...,\phi_{M})$. The
multimode quadrature operator is given by~\cite{VogelBook}:
\begin{equation}
\hat{x}_{\bm{\phi}}=\sum_{m=1}^{M}c_{m}\;\hat{x}_{m}(\phi_{m})
\label{N11}
\end{equation}
is a sum of single-mode phase-rotated quadratures
\begin{equation}
\hat{x}_{m}(\phi_{m})=\hat{a}_{m}\exp(i\phi_{m})
+\hat{a}_{m}^{\dagger}\exp(-i\phi_{m}).
\label{N12}
\end{equation}
The truncated nonclassicality witness $\tilde{S}_{x_{\phi}}$,
given by Eq.~(\ref{SxPhiBar}), can also be used in a
single-mode case. The $\bm{\phi}$-optimized quadrature
squeezing is referred to as \emph{principal} squeezing and is
defined by the witness~\cite{Luks88,Miran10}:
\begin{equation}
S_{{\rm opt}}=\min_{\bm{\phi}}S_{x_{\phi}}\ncl0, \label{Sopt}
\end{equation}
or the truncated witness
\begin{equation}
\tilde{S}_{{\rm opt}}=\max(0,-S_{{\rm opt}}-S_{0})
=\max_{\bm{\phi}}\tilde{S}_{x_{\phi}}\nclg0. \label{SoptBar}
\end{equation}
Note that all entanglement witnesses are also nonclassicality
witnesses, but not vice versa. An example of the single-mode
evolution exhibiting the SV and SR of the nonclassicality
witnesses, corresponding to the quadrature and principal
squeezing, is shown in Fig.~1 for the anharmonic model
described in Sec.~VI.

Explicit examples of many other two- and multimode nonclassicality
witnesses, corresponding to violations of classical inequalities,
can be found in, e.g.,
Refs.~\cite{Reid86,Perina91,Korbicz,VogelBook,Miran10,Wang10,Miran99}.

\section{Entanglement witnesses}

An effective method of constructing entanglement witnesses can be
based on the Shchukin-Vogel entanglement criterion~\cite{Shchukin}
(or its generalizations~\cite{Miran09}) for distinguishing states
with positive partial transposition from those with nonpositive
partial transposition (NPT).

In analogy to the matrices of normally-ordered moments $M^{\rm
(n)}_{\hat f}(\hat\rho)$, given by Eq.~(\ref{w2}), one can define
the following matrix of partially-transposed moments:
\begin{eqnarray}
 M_{\hat f}(\hat\rho^\Gamma)
= \Mat{ \pt{\hat f_{1}^\dagger \hat f_{1}} & \pt{\hat
f_{1}^\dagger \hat f_{2}} & \cdots & \pt{\hat f_{1}^\dagger \hat
f_{N}} } { \pt{\hat f_{2}^\dagger \hat f_{1}} & \pt{\hat
f_{2}^\dagger \hat f_{2}} & \cdots & \pt{\hat f_{2}^\dagger \hat
f_{N}} } { \vdots & \vdots & \ddots & \vdots } { \pt{\hat
f_{N}^\dagger \hat f_{1}} & \pt{\hat f_{N}^\dagger \hat f_{2}} &
\cdots & \pt{\hat f_{N}^\dagger \hat f_{N}} },
  \label{w3}
\end{eqnarray}
where $\hat f = \sum_{i}^N c_{i} \hat f_{i}$ for arbitrary
complex numbers $c_{i}$, $\langle\hat f_{i}^\dagger \hat
f_{j}\rangle^{\Gamma}\equiv {\rm tr} (\hat f_{i}^\dagger \hat
f_{j} \hat\rho^\Gamma)$ and $\Gamma$ denotes partial
transposition. The Shchukin-Vogel entanglement
criterion~\cite{Shchukin,Miran09} can be written as:
\begin{criterion} 
A bipartite state $\hat\rho$ is NPT if and only if there exists
${\hat f}$, such that $\det[M_{\hat f}(\hat\rho^\Gamma)]$ is
negative.
\end{criterion}
This criterion resembles Criterion~2 of the nonclassicality.
Thus, analogously to the nonclassicality witnesses, we refer
to such matrices $M_{\hat f}(\hat\rho^\Gamma)$ of partially
transposed moments and their functions (like determinants) as
(state-dependent nonlinear) \emph{entanglement witnesses}. It
is worth noting that according to the original definition,
entanglement witnesses correspond to observables rather than
expectation values~\cite{Horodecki96}: An entanglement
witness is a Hermitian operator $\hat W$ such that ${\rm
tr}(\hat W \hat \rho_{\rm sep})\ge 0$ for all separable
states $\hat\rho_{\rm sep}$, while ${\rm tr}(\hat W
\hat\rho_{\rm ent})<0$ for some entangled states
$\hat\rho_{\rm ent}$. This concept was later generalized to
nonlinear entanglement witnesses~\cite{Hyllus,Guhne}.
Although our usage of the term entanglement witness differs
slightly from the original usage, we believe that it can
improve readability of our paper, while keeping unchanged the
main idea of entanglement witnesses.

Here we give only two examples of such entanglement witnesses
based on Criterion~3. Let us apply the following
Hillery-Zubairy classical inequalities~\cite{Hillery}:
\begin{eqnarray}
\langle\hat{n}_{1}\hat{n}_{2}\rangle\cl|\langle\hat{a}_{1}
\hat{a}_{2}^{\dagger}
\rangle|^{2},\quad\langle\hat{n}_{1}\rangle\langle\hat{n}_{2}
\rangle\cl|\langle\hat{a}_{1}\hat{a}_{2}\rangle|^{2},
\label{Hillery12}
\end{eqnarray}
where $\hat{n}_{i}=\hat{a}_{i}^{\dagger}\hat{a}_{i}$ is the photon
number operator, and $\hat{a}_{i}$ ($\hat{a}_{i}^{\dagger}$) is
the annihilation (creation) operator for mode $i=1,2$. Thus, we
can define the following truncated witnesses
\begin{eqnarray}
\tilde{H} & = & \max(0,|\langle\hat{a}_{1}\hat{a}_{2}^{\dagger}
\rangle|^{2}-\langle\hat{n}_{1}\hat{n}_{2}\rangle)\ent0,
\label{Hillery1}\\
\tilde{H}' & = &
\max(0,|\langle\hat{a}_{1}\hat{a}_{2}\rangle|^{2}-
\langle\hat{n}_{1}\rangle\langle\hat{n}_{2}\rangle)\ent0.
\label{Hillery2}
\end{eqnarray}
which can be positive only for some \emph{entangled} states, as
marked by the symbol $\ent$. These inequalities can be derived in
various ways, e.g., from the Cauchy-Schwarz
inequality~\cite{Hillery} or from entanglement criteria based on
partial transposition~\cite{Shchukin,Miran09}. Thus, $\tilde{H}$
and $\tilde{H}'$ are entanglement witnesses, so the SV of the
concurrence implies also the SV of $\tilde{H}(t)$ and
$\tilde{H}'(t)$ (if they were nonzero for some evolution times).
It is worth noting that the inequalities in Eq.~(\ref{Hillery12})
are satisfied not only by separable states but also by all
classical states (marked by $\cl$) since they can be derived from
nonclassicality criteria based on the $P$ function~\cite{Miran10}.

Another simple choice of an entanglement witness can be
related, e.g., to the violation of Bell's inequality. For
two-qubit states, a degree of violation of Bell's inequality,
in its version due to Clauser, Horne, Shimony, and Holt
(CHSH)~\cite{Clauser}, can be defined
as~\cite{Horodecki95,Miran04pla}:
\begin{eqnarray}
B^{2}(\hat{\rho})\equiv\max\,\Big[0,\,\max_{j<k}\;(u_{j}+u_{k})-1\,\Big],
\label{nonlocality}
\end{eqnarray}
where $u_{j}$ $(\, j=1,2,3)$ are the eigenvalues of
$U_{\hat{\rho}} =T_{\hat{\rho}}^{T}\, T_{\hat{\rho}}$,
$T_{\hat{\rho}}$ is a real matrix with elements $t_{ij}={\rm
Tr}\,[\hat{\rho}\,(\hat{\sigma}_{i}\otimes\hat{\sigma}_{j})]$,
and $\hat{\sigma}_{j}$ are Pauli's spin matrices. For
brevity, although not precisely, $B$ is often referred to as
a \emph{nonlocality} (measure).  Analogously to the
concurrence $C$, the nonlocality $B$ is defined as the
maximum of zero and another quantity, which implies that it
is possible to observe the SV and SR of $B(t)$ in a dynamical
scenario.

If a two-qubit state $\hat{\rho}$ violates Bell's inequality then
it is also entangled, but not vice versa, i.e., there are mixed
states $\hat{\rho}$ (e.g., Werner's states discussed below), for
which $C(\hat{\rho})>0$ and $B(\hat{\rho})=0$. Thus,
$B(\hat{\rho})$ can be considered as an \emph{entanglement
witness}. The SV of an entanglement measure implies the SV of an
nonlocality measure (if the latter was nonzero at some evolution
time). Note that for two-qubit pure states
$B(\hat{\rho})=C(\hat{\rho})$, so in this case the nonlocality is
not only an entanglement witness but also an entanglement measure.

\section{Sudden decays of nonclassicality witnesses
for noninteracting modes}

Let us first give a simple example of the environment-induced
sudden vanishing of the entanglement that is closely related
to the original idea of finite-time sudden decays. As a
generalization, we also study sudden vanishings of several
other nonclassicality witnesses, which occur at times
different than those for the entanglement vanishing.

By contrast to the following sections, we analyze the
entanglement of two modes (qubits), which are not directly
interacting with each other but only with independent
reservoirs. Specifically, we describe the SV of the
nonclassicality of initially entangled states, due to
interaction with the reservoirs under Markov's approximation,
by applying the standard master equation for the reduced
density operator $\hat{\rho}$:
\begin{eqnarray}
\frac{\partial}{\partial t}\hat{\rho}
=\sum_{k=1,2}\frac{\gamma_{k}}{2}[\bar{n}_{k}
(2\hat{a}_{k}^{\dagger}\hat{\rho}_{k}\hat{a}_{k}
-\hat{a}_{k}\hat{a}_{k}^{\dagger}\hat{\rho}
-\hat{\rho}\hat{a_{k}}\hat{a}_{k}^{\dagger})
\quad\quad\quad\label{ME}\\
+(\bar{n}_{k}+1) (2\hat{a}_{k}\hat{\rho}\hat{a}_{k}^{\dagger}
-\hat{a}_{k}^{\dagger}\hat{a}_{k}\hat{\rho}
-\hat{\rho}\hat{a}_{k}^{\dagger}\hat{a}_{k})]
-\frac{i}{\hbar}[\hat{\cal H}_{S},\hat{\rho}],\nonumber
\end{eqnarray}
where $\gamma_{k}$ are the damping rates, $\bar{n}_{k}$ are
the mean thermal photon numbers,
$\bar{n}_{k}=\{\exp[\hbar\omega_{k}/(k_{B}T)]-1\}^{-1}$, $T$
is the reservoirs temperature at thermal equilibrium, and
$k_{B}$ is Boltzmann's constant. We assume the reservoirs to
be at zero temperature, so we set
$\bar{n}_{1}=\bar{n}_{2}=0$. The Hamiltonian $\hat{\cal
H}_{S}$ is just the sum of free Hamiltonians for the two
noninteracting system modes. We solve the master equation by
applying the Monte Carlo wave function simulation with the
collapse operators
$\hat{c}_{1k}=\sqrt{\gamma(1+\bar{n}_{k})}\hat{a}_{k}$ and
$\hat{c}_{2k}=\sqrt{\gamma\bar{n}_{k}}\hat{a}_{k}^{\dagger}$~\cite{Tan}.

It is worth noting that from the standard physical point of
view, the quantum entanglement between two systems, and the
related violation of Bell's inequalities, can be considered
if the systems are spatially separated and are physically
uncoupled~\cite{Wojcik}. It is seen that this model (contrary
to the models studied in the following section) satisfies the
second condition.

Our example of the environment-induced sudden vanishing of
quantumness and nonlocality is provided for a system coupled to
two independent reservoirs. It is worth mentioning that common
reservoirs in some cases can also enhance entanglement both for
two qubits and two modes. This is possible due to a mixing
mechanism rather than an induced interaction among
them~\cite{Benatti}.

Let us analyze the decoherence of the initial Werner-like
state defined as~\cite{Miran04pla}:
\begin{equation}
\hat{\rho}_{m}(0)=p|\Psi_{m}\rangle\langle\Psi_{m}|
+\frac{1-p}{4}\hat{I}, \label{werner}
\end{equation}
for $0\le p\le1$, $m=1$, and
$|\Psi_{1}\rangle=(|00\rangle+|11\rangle)/\sqrt{2}$. Here,
$\hat{I}$ is the identity operator. Under this initial
condition, the solution of the master equation can be given
in the standard computational basis as~\cite{Miran04pla}:
\begin{equation}
\hat{\rho}_{1}(t)=\frac{1}{4}\left[\begin{array}{cccc}
h^{(+)} & 0 & 0 & 2p\sqrt{g_{1}g_{2}}\\
0 & h_{1}^{(+)} & 0 & 0\\
0 & 0 & h_{2}^{(+)} & 0\\
2p\sqrt{g_{1}g_{2}} & 0 & 0 &
(1+p)g_{1}g_{2}\end{array}\right],\label{rho1}
\end{equation}
where $h^{(+)}=(2-g_{1})(2-g_{2})+pg_{1}g_{2}$,
$h_{k}^{(+)}=g_{3-k}[2-(1+p)g_{k}]$, and
$g_{k}=\exp(-\gamma_{k}t)$ for $k=1,2$. The concurrence and
nonlocality decay as follows~\cite{Miran04pla}:
\begin{eqnarray}
C(t)=\max\left\{ 0,\frac{1}{2}\sqrt{g_{1}g_{2}}\Big(2p\right.\hspace{3cm}\nonumber \\
\left.-\sqrt{[2-(1+p)g_{1}][2-(1+p)g_{2}]}\Big)\right\}
,\label{C1}
\end{eqnarray}
\begin{equation}
B^{2}(t)=\max\left(0,2p^{2}g_{1}g_{2}-1\right),\label{B1}
\end{equation}
respectively. For comparison, we also calculate the decays of
the two witnesses of the photon-number-difference
correlations:
\begin{equation}
\tilde{S}(t)=\max\left[0,\frac{1}{4}(g_{1}^{2}+g_{2}^{2}+2pg_{1}g_{2})
-S_{0}\right],\label{S1}
\end{equation}
\begin{align}
\tilde{D}(t)=\max\left[0,\frac{1}{2}g_{1}g_{2}(1+p)-D_{0}^{2}
-D_{0}(g_{1}-g_{2})\right].\label{D1}
\end{align}
For simplicity, let us assume now the same reservoir damping
rate $\gamma$, so $g_{1}=g_{2}\equiv g$. Then, the SV times
for the above entanglement and nonclassicality witnesses can
be different from each other as they are given by
\begin{equation}
t_{{\rm SV}}^{(C)}=\frac{1}{\gamma}\ln
\left(\frac{1+p}{2(1-p)}\right),\label{t_c1}
\end{equation}
\begin{equation}
t_{{\rm
SV}}^{(B)}=\frac{1}{\gamma}\ln\left(\sqrt{2}p\right),\label{t_b1}
\end{equation}
\begin{equation}
t_{\rm SV}^{(\tilde{S})}
=\frac{1}{2\gamma}\ln\left(\frac{1+p}{2S_{0}}\right),\label{t_s1}
\end{equation}
\begin{equation}
t_{{\rm SV}}^{(\tilde{D})}=
\frac{1}{2\gamma}\ln\left(\frac{1+p}{2
D_{0}^{2}}\right).
\label{t_d1}
\end{equation}
The results are shown in Fig.~2 assuming some specific values
of the damping constant $\gamma$ and the initial Werner state
$\hat{\rho}_{1}(0)$ with parameter $p$.

\begin{figure}
\centerline{\epsfxsize=8.5cm\epsfbox{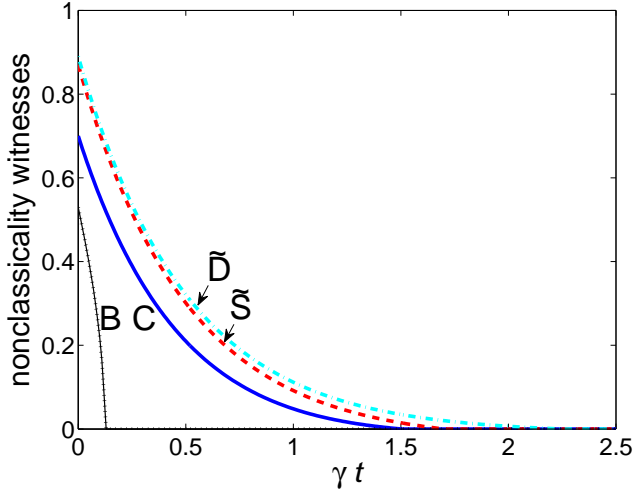}}
\caption{(Color online) An example of the environment-induced
sudden vanishing of the nonclassicality witnesses for two
\emph{noninteracting} modes. The damping model is described
in Sec. IV for the initial Werner-like state $\rho_{1}$ with
$p=0.8$. Key: the concurrence $C$ (solid curve), nonlocality
$B$ (dotted curve), and two witnesses describing the
photon-number-difference correlations: $\tilde S$ (dashed
curve) for $S_0=0.03$ and $\tilde D$ (dot-dashed curve) for
$D_0=0.1$.}
\end{figure}

In conclusion, we have given a simple example of the decaying
entanglement between two qubits, which are not directly
interacting with each other, but they are only coupled to the
environment. We have observed the SVs of the two
nonclassicality witnesses, which are different from the SVs
of the entanglement and nonlocality measures.

\section{Periodic sudden vanishing of nonclassicality
witnesses of interacting modes}

\subsection{Frequency conversion model}

Here we give an illustrative example of \emph{periodic} sudden
vanishing of nonclassicality witnesses during a unitary evolution
of two interacting modes. This is in contrast to the standard
analysis of sudden decays applied solely to dissipative systems.
Note that one can easily include the dissipation (as studied,
e.g., in the former section) to observe the proper finite-time
sudden decays and SRs analogous to the standard ones.

As a simple model to study SV and SR, let us study the parametric
frequency conversion described by the interaction Hamiltonian
\begin{eqnarray}
\hat{\cal H}=\hbar\kappa[\hat{a}_{1}^{\dagger}\hat{a}_{2}
\exp(-i\Delta\omega
t)+\hat{a}_{1}\hat{a}_{2}^{\dagger}\exp(i\Delta\omega
t)],\label{N01}
\end{eqnarray}
which is a prototype Hamiltonian describing two
linearly-coupled harmonic oscillators. It can be applied to a
variety of physical phenomena including the process of
exchanging photons between two optical fields of different
frequencies: a signal mode with frequency $\omega_{1}$ and an
idler mode with frequency $\omega_{2}$. Then $\hat{a}_{1}$
and $\hat{a}_{2}$ are the annihilation operators for the
signal and idler modes, respectively, and $\kappa$ is the
real coupling constant. For simplicity, we assume a resonant
case $\Delta\omega=\omega+\omega_{2}-\omega_{1}$.

The well-known solutions of the Heisenberg equations of motion for
the signal, $\hat{b}_{1}(t)$, and idler, $\hat{b}_{2}(t)$, modes
are given by~\cite{Louisell64}:
\begin{eqnarray}
\hat{b}_{1}(t) & = & \hat{a}_{1}\cos(\kappa t)\,-i\,\hat{a}_{2}\sin(\kappa t),\nonumber \\
\hat{b}_{2}(t) & = & \hat{a}_{2}\cos(\kappa
t)\,-i\hat{a}_{1}\sin(\kappa t).\label{N02}
\end{eqnarray}
The corresponding solution of the Schr\"odinger equation is
\begin{equation}
|\psi(t)\rangle=\sum_{n_{1},n_{2}}c_{n_{1},n_{2}}
\frac{[\hat{b}_{1}^{\dagger}(-t)]^{n_{1}}}
{\sqrt{n_{1}!}}\frac{[\hat{b}_{2}^{\dagger}(-t)]^{n_{2}}}
{\sqrt{n_{2}!}}|00\rangle \label{N02a}
\end{equation}
assuming that the system is initially in a superposition of Fock
states,
$|\psi(0)\rangle=\sum_{n_{1},n_{2}}c_{n_{1},n_{2}}|n_{1},n_{2}\rangle.$
The total number of photons is a constant of motion,
$\hat{n}_{1}(t)+\hat{n}_{2}(t)=$const.

An important property of the (undamped) parametric frequency model
is that the nonclassicality of an arbitrary state is unchanged
during its evolution. By applying the results of
Refs.~\cite{Glauber,Mista,Miran98}, one can find that the time
evolution of the QPD for the frequency-converter model, described
by Eq.~(\ref{N01}), with arbitrary initial fields is simply given
by
\begin{equation} {\cal W}^{(s)}(\alpha_{1,}\alpha_{2},t)={\cal
W}^{(s)}\left[\beta_{1}(\alpha_{1},\alpha_{2},-t),\beta_{2}(\alpha_{1},\alpha_{2},-t),0\right],\label{QPD}
\end{equation}
where $\beta_{1,2}(\alpha_{1},\alpha_{2},t)$ are the solutions of
the corresponding \emph{classical} equations of motion for the
frequency conversion model:
\begin{eqnarray}
\beta_{1}(\alpha_{1},\alpha_{2},t) & = & \alpha_{1}\cos(\kappa t)-i\alpha_{2}\sin(\kappa t),\nonumber \\
\beta_{2}(\alpha_{1},\alpha_{2},t) & = & \alpha_{2}\cos(\kappa
t)-i\alpha_{1}\sin(\kappa t).\label{N03}
\end{eqnarray}
Equation~(\ref{QPD}) means that the two-mode QPD for the model
discussed is constant along classical trajectories. Thus, if the
initial fields are nonclassical, their degree of nonclassicality
(as defined, e.g., in Refs.~\cite{Lee91,Lutkenhaus,Kenfack})
remains unchanged at any evolution times of the system. But yet we
can observe SV and SR of entanglement and nonclassicality
witnesses as will be shown in the following subsections.

\subsection{Evolution of a pure state}

\begin{figure}
\epsfxsize=8cm\epsfbox{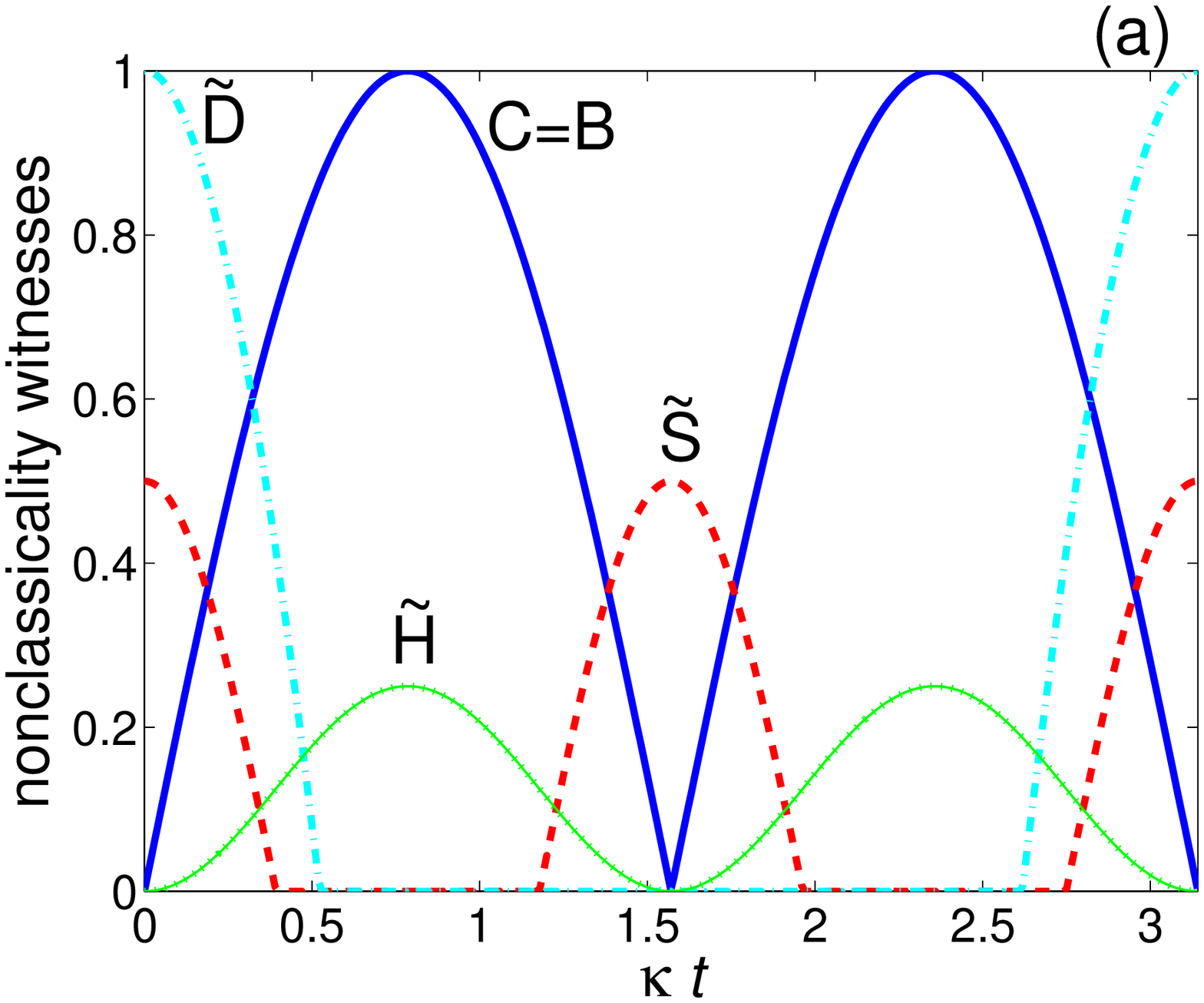}

\vspace*{-2mm} \epsfxsize=8cm\epsfbox{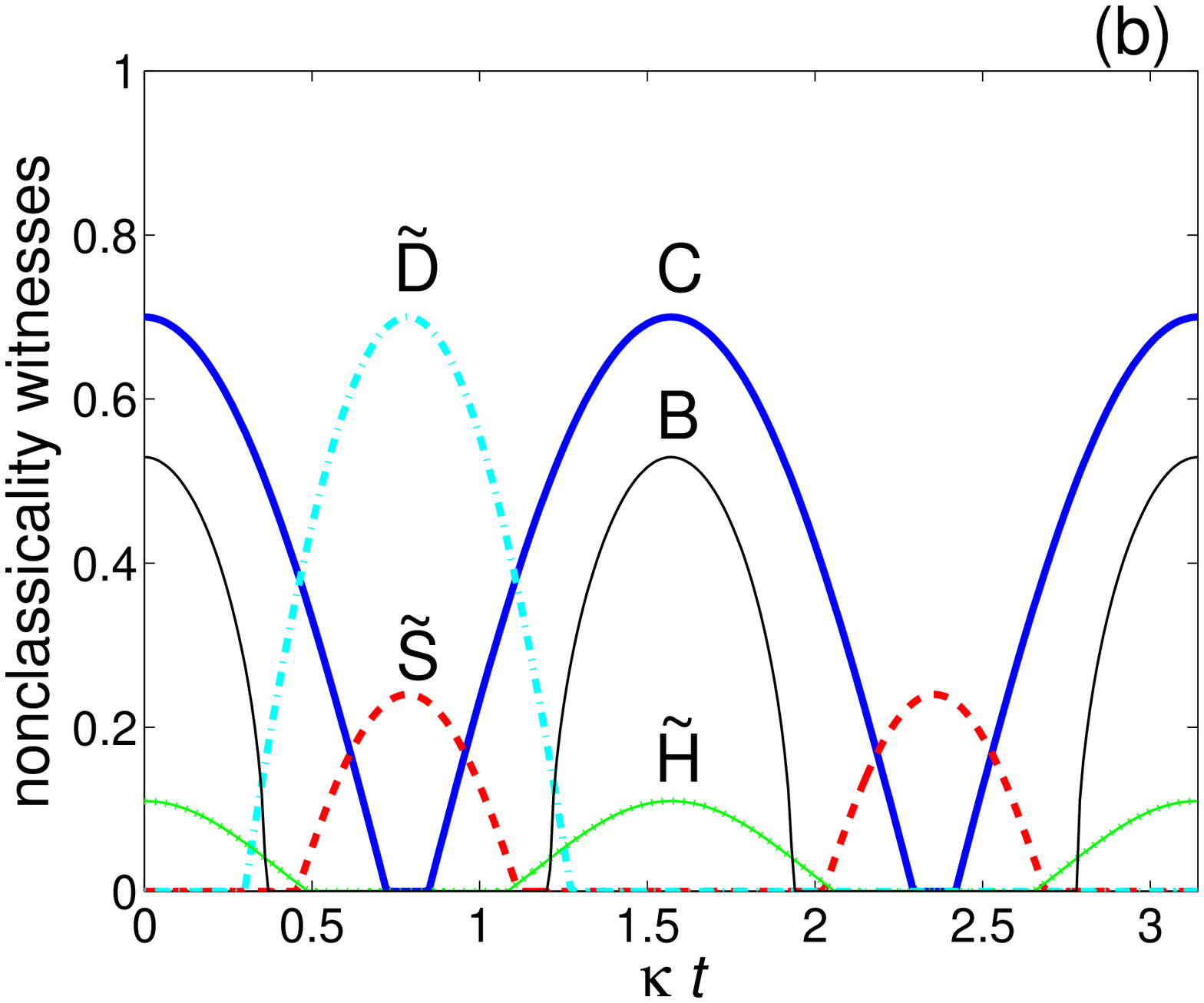}
 \caption{(Color online) Simple examples of the sudden vanishing and
reappearance of the concurrence and other truncated
nonclassicality witnesses for two interacting modes. The
unitary evolution of the frequency model is shown assuming:
(a) the initial pure state $|01\rangle$ discussed in Sec.~V~B
and (b) the initial mixed state, given by Eq.~(\ref{m2b})
with $p=0.8$, analyzed in Sec.~V~C. Key: $C$ (thick solid
curve) is the concurrence, $B$ (thin solid curve) is the
nonlocality; $\tilde H$ (dotted curve) is the entanglement
witness, given by Eq.~(\ref{Hillery1}), describing the
violation of the first Hillery-Zubairy inequality; $\tilde S$
(dashed curve) for $S_0=1/2$ and $\tilde D$ (dot-dashed
curve) for $D_0=1$ are nonclassicality witnesses describing
the photon-number-difference correlations, which are given by
Eqs.~(\ref{Sn}) and~(\ref{Dn}), respectively. From the
standard point of view, a SR should appear only after some
finite evolution time after the occurrence of the preceding
SV. It is seen that this condition is satisfied for all the
witnesses of the mixed-state evolution (b), but only for some
witnesses of the pure-state evolution (a).}
\end{figure}

Let us first analyze the parametric frequency conversion for the
initial state $|\psi(0)\rangle=|01\rangle$. The system evolves,
according to Eq.~(\ref{N02a}), into
\begin{equation}
|\psi(t)\rangle=\cos(\kappa t)|01\rangle-i\sin(\kappa
t)|10\rangle.\label{N21}
\end{equation}
It is a nonclassical state described by the following
\emph{singular} (so negative) \emph{$P$} function
\begin{eqnarray}
P(\alpha_{1},\alpha_{2},t)=\delta[\beta_{1}(\alpha_{1},\alpha_{2},t)]
\left(1+\frac{\partial}{\partial\beta_{2}(\alpha_{1},\alpha_{2},t)}\right.\nonumber \\
\left.\times\frac{\partial}{\partial\beta_{2}^{*}
(\alpha_{1},\alpha_{2},t)}\right)
\delta[\beta_{2}(\alpha_{1},\alpha_{2},t)],\quad\label{N22}
\end{eqnarray}
which is given in terms of Dirac's $\delta$ function, its
derivative, and the solutions of the classical equations of
motion, given by Eq.~(\ref{N03}). Elementary calculations
lead to the following expressions for the concurrence and
nonlocality
\begin{equation}
C(t)=B(t)=|\sin(2\kappa t)|,\label{e1}
\end{equation}
the entanglement witness describing the violation of the
first Hillery-Zubairy inequality
\begin{equation}
\tilde{H}(t)=\frac{1}{4}\sin^{2}(2\kappa t),\label{e4}
\end{equation}
and the nonclassicality witnesses for the
photon-number-difference correlations
\begin{eqnarray}
\tilde{S}(t)&=&\max\left[0,\cos^{2}(2\kappa t)
-S_{0}\right],\label{e2}
\\
\tilde{D}(t)&=&\max\left\{0,D_{0}[2\cos(2\kappa t)-D_{0}]\right\}.
\label{e3}
\end{eqnarray}
Analogously, one also finds the photon-number sub-Poisson
statistics of the fields as described by the modified Mandel
parameters $\tilde{Q}_{1}=\sin^{2}(\kappa t)$ and
$\tilde{Q}_{2}=\cos^{2}(\kappa t)$. All these nonclassical
witnesses exhibit periodic SV and SR effects as shown in
Fig.~3(a). For example, the SV and SR of the concurrence
corresponds to the maximum value of $\tilde{S}$. Analogously,
we could observe the out-of-phase SVs and SRs of Mandel's $Q$
parameters, which can be clearly understood by recalling the
classical-like interpretation of two linearly coupled
oscillators when one of them is initially excited ($Q_{2}>0$)
and the other is unexcited ($Q_{1}=0$). During the evolution,
the excitation is transferred periodically between the
oscillators.

One can raise an objection concerning the above example that
a SV of the concurrence is instantly followed by a SR, so
they are not the proper SV and SR effects. The same behavior
is found for the other witnesses including $\tilde{D}$ for
$D_0=0$, and $\tilde{S}$ for $S_0=0$. From the more standard,
or more orthodox, point of view, a SV (of some witness)
should not be instantly followed by a SR. By contrast, the SV
times differ from the SR times for $\tilde{D}$ with $D_0>0$
and for $\tilde{S}$ with $S_0>0$ [as shown in Fig.~3(a)] that
is required in the orthodox approach.

Other, even more convincing, examples of the SV and SR
effects can be found by analyzing the evolution of initially
mixed states as will be shown below.

\subsection{Evolution of a mixed state}

Let us choose the initial state to be a Werner-like state
$\hat{\rho}_{0}(0)$, given by Eq.~(\ref{werner}) for $m=0$
and $|\Psi_{0}\rangle=(|01\rangle-i|10\rangle)/\sqrt{2}$.
This state evolves as follows
\begin{equation}
\hat{\rho}_{0}(t)=p|\Psi_{0}(t)\rangle\langle\Psi_{0}(t)|+\frac{1-p}{4}\hat{I},
\label{m2b}
\end{equation}
where
\begin{equation}
|\Psi_{0}(t)\rangle=
\frac{1}{\sqrt{2}}\big[f_{-}(t)|01\rangle-if_{+}(t)|10\rangle\big]
\end{equation}
with $f_{\pm}(t)=\cos(\kappa t)\pm\sin(\kappa t)$. We find the
following evolutions of the entanglement witnesses and the
corresponding times of the first SV:
\begin{equation}
C(t)=\max[0,p|c|-(1-p)/2]\;\Rightarrow\; t_{{\rm
SV}}^{(C)}=f\left(\frac{1-p}{2p}\right),\label{m3}
\end{equation}
\begin{equation}
B^{2}(t)=\max[0,p^{2}\left(1+c^{2}\right)-1]\;\Rightarrow\;
t_{{\rm
SV}}^{(B)}=f\left(\frac{\sqrt{1-p^{2}}}{p}\right),\label{m4}
\end{equation}
\begin{equation}
\tilde{H}(t)=\frac{1}{4}\max[0,(pc)^{2}-(1-p)]\;\Rightarrow\;
t_{{\rm
SV}}^{(\tilde{H})}=f\left(\frac{\sqrt{1-p}}{p}\right),\label{m5}
\end{equation}
where $f(x)=\arccos x/(2\kappa)$ and $c=\cos(2\kappa t)$. The
first SR occurs at the time
\begin{equation}
\kappa t_{{\rm
SR}}^{(i)}=\pi/2-\kappa t_{{\rm SV}}^{(i)}\label{t_SR}
\end{equation}
for $i=C,B,\tilde{H}$. It is seen in Fig.~3(b) for $p=0.8$
that the first SVs and SRs occur in the following order:
\begin{equation}
t_{{\rm SV}}^{(B)}<t_{{\rm SV}}^{(\tilde{H})}<t_{{\rm
SV}}^{(C)}\;\Rightarrow\; t_{{SR}}^{(B)}>t_{{\rm
SR}}^{(\tilde{H})}>t_{{\rm SR}}^{(C)}.
\end{equation}
On the other hand, the nonclassicality witnesses $\tilde{D}$ and
$\tilde{S}$, given by Eqs.~(\ref{Sn}) and~(\ref{Dn}),
respectively, evolve as
\begin{eqnarray}
\tilde{S}(t)  =  \max[0,(1-p)/2+p^{2}\sin^{2}(2\kappa
t)-S_{0}],\label{m7}
\\
\tilde{D}(t)  = \max[0,(1-p)/2+2D_{0}p\sin(2\kappa
t)-D_{0}^{2}].\label{m7b}
\end{eqnarray}
For $S_{0}=0$ and $p<1,$ we do not observe a complete
vanishing of $\tilde{S}(t)$. For $S_{0}=0$ and $p=1$ (which
corresponds to the initial Bell state), $\tilde{S}(t)$
periodically vanishes to zero and instantly increases, so it
is not a good example of the SV and SR effects. However, for
$0<p<1$ we can observe the proper SV and SR effects as shown
in Fig.~3(b). The first SVs occur at the times
\begin{eqnarray}
t_{{\rm SV}}^{(\tilde{S})} & = & \frac{\pi}{4\kappa}+f\left(
\frac{\sqrt{2S_{0}+p-1}}{\sqrt{2}p} \right),\label{m8}
\\
t_{{\rm SV}}^{(\tilde{D})}&=&\frac{\pi}{4\kappa}
+f\left(\frac{2D_{0}^{2}+p-1}{4D_{0}p}\right),\label{m8a}
\end{eqnarray}
and the first SRs occur at $t_{{\rm
SR}}^{(\tilde{S})}=\pi/\kappa - t_{{\rm SV}}^{(\tilde{S})}$
and $t_{{\rm SR}}^{(\tilde{D})}=3\pi/(2\kappa) - t_{{\rm
SV}}^{(\tilde{D})}$. Note that the first appearances of these
witnesses occur at earlier times, i.e., $t=\pi/(2\kappa) -
t_{{\rm SV}}^{(i)}$ for $i=\tilde{S},\tilde{D}$. It is seen
that we can always choose threshold values $S_{0}$ and
$D_{0}$ for any $0<p<1$ in such a way to observe the SVs and
SRs of these witnesses for the photon-number-difference
correlations at arbitrary evolution times also when the
system is disentangled.

\section{Periodic sudden vanishing of nonclassicality witnesses
for a single mode}

Finally, let us analyze a single-mode anharmonic oscillator
described by the interaction Hamiltonian
\begin{eqnarray}
\hat{\cal H} & = &
\frac{1}{2}\hbar\kappa(\hat{a}^{\dagger})^{2}\hat{a}^{2}.\label{a1}
\end{eqnarray}
This is a prototype model of various fundamental phenomena
including the optical Kerr effect. For simplicity, here we refer
to this effect only. Under this interaction, the initial coherent
state $|\alpha_{0}\rangle$ evolves periodically into a
nonclassical state
\begin{equation}
|\psi(t)\rangle=e^{-|\alpha_{0}|^{2}/2}\sum_{n=0}^{\infty}\frac{\alpha_{0}^{n}}{\sqrt{n!}}\exp\left[\frac{i}{2}n(n-1)\tau\right]|n\rangle,\label{a2}
\end{equation}
where $\tau$ is a rescaled time $\kappa t$. It is worth
noting that the Kerr state, given by Eq.~(\ref{a2}), becomes
at some evolution times a superposition of macroscopically
distinguishable two~\cite{Yurke86} or more~\cite{Miran90}
coherent states, which are often referred to as the
Schr\"odinger cat and kitten states, respectively. Among many
nonclassical intriguing properties of the model (see, e.g.,
Ref.~\cite{Tanas03} and references therein), the Kerr state
exhibits high-degree quadrature
squeezing~\cite{Luks88,Miran91}. We find that the single-mode
normally-ordered variance $S_{x_{\phi}}$ of the quadrature
operator
$\hat{x}_{\phi}=\hat{x}_1(\phi)=\hat{a}\exp(-i\phi)+\hat{a}^{\dagger}\exp(i\phi)$
can be compactly written as follows:
\begin{equation}
S_{x_{\phi}}  = 2|\alpha_{0}|^{2}[1+f_{12}\cos(\tau_{12}+\tau)
-f_{21}(\cos\tau_{21}+1)] \label{SxPhiAOM}
\end{equation}
in terms of the auxiliary functions defined by $\tau_{kl} =
k|\alpha_{0}|^{2}\sin(l\tau)+2(\phi-\phi_{0})$ and $f_{kl}  =
\exp\{k|\alpha_{0}|^{2}[\cos(l\tau)-1]\}$ with
$\alpha_{0}=|\alpha_{0}|\exp(i\phi_{0})$. Quadrature squeezing
occurs if $S_{x_{\phi}}\ncl 0$ or, equivalently, if the truncated
witness $\tilde{S}_{x_{\phi}}\nclg 0$, defined by
Eq.~(\ref{SxPhiBar}) with Eq.~(\ref{SxPhiAOM}) and ${\bm
\phi}=\phi$. For simplicity, we set a threshold value $S_{0}$ to
be zero in this section and in Fig.~1. By applying the results of
Refs.~\cite{Luks88,Miran91}, we can compactly write the
$\phi$-optimized variance $S_{\rm opt}$ describing the principal
squeezing as follows
\begin{equation}
S_{\rm opt}(t)  =  2|\alpha_{0}|^{2} \Big(1-f_{21}
-\sqrt{f_{22}+f_{41}-2f_{12}f_{21}\cos\tau'}\Big),\label{SoptAOM}
\end{equation}
where $\tau'=\tau_{12}-\tau_{21}+\tau$. Analogously to the
former squeezing criteria, the principal squeezing occurs if
$S_{\rm opt}\ncl 0$ or if the truncated witness
$\tilde{S}_{\rm opt}\nclg 0$, as given by
Eqs.~(\ref{SoptBar}) and~(\ref{SoptAOM}). Our results are
presented in Fig.~1 for some specific amplitude of the
initial coherent state.

Note that the periodic vanishing of the entanglement and
nonclassicality witnesses, analyzed here and in Sec.~V,
should not be confused with the oscillations of the
entanglement measures in systems interacting with
non-Markovian reservoirs (see, e.g., Ref.~\cite{Mazzola}).
The SV and SR effects in such systems have different
character than studied here. Mazzola \emph{et
al.}~\cite{Mazzola} observed the oscillations in short times,
which disappear after some finite time and are related to the
non-Markovian character of the reservoirs. In contrast, in
the examples presented here, the periodic behavior of the
nonclassicality witnesses persists as being related to the
unitary evolution of the states.

It is worth stressing again that the aperiodic SV and SR effects,
which are analogous to the typical sudden decays of the
entanglement, can be observed by inclusion of the dissipation.
Assuming Markov's approximation, one can apply the master
equation, given by Eq.~(\ref{ME}) in a special case for a single
mode ($k=1$). Then the SVs and SRs become aperiodic and the final
SV occurs after some evolution time, which depends on the
dissipation. However, the dissipation is not a necessary condition
for the SV occurrence in this model.

The SV and SR of the entanglement in two-mode dissipative
coupled Kerr models was studied in Ref.~\cite{KerrCoupler}.
Here we showed that the periodic SV and SR of squeezing can
be observed even in the single-mode nondissipative Kerr
model. This example confirms our conclusion of the general
occurrence of the SV and SR of nonclassicality witnesses even
for single-mode undamped systems.

\section{Conclusions}

We have applied the concepts of the SV and SR of quantum
entanglement measures to study the SV and SR of entanglement and
nonclassicality witnesses.

Our main observations can be summarized as follows:

(i) SVs can be encountered not only in the dissipation of
entanglement but also of other nonclassical correlation
parameters, related to violations of classical
inequalities~\cite{Miran10,VogelBook}.

(ii) SVs occur not only in the dissipation of bipartite or
multipartite (multimode) interacting or noninteracting systems but
also in a single-qubit or single-mode systems. Our examples
include single-mode squeezing of photon number, squeezing of
quadrature operators~\cite{VogelBook}, and violations of other
classical inequalities~\cite{Miran10}.

(iii) Non-dissipative systems, which are initially even in pure
states, can also exhibit periodic SVs of nonclassical phenomena
and the related nonclassicality witnesses. For instance, the
quadrature squeezing of light in a Kerr medium exhibits periodic
SVs for some finite periods of time. In order to observe the
proper finite-time sudden decays analogous to the standard sudden
decays of entanglement~\cite{Yu04}, one should add dissipation by
coupling such systems to the environment. The damping causes
irregularity and loss of periodicity of the evolution of the
nonclassicality witnesses. We can conclude that the damping
accelerates the occurrence of the first SVs but it is not a
necessary condition for their occurrence.

With the help of the nonclassicality
criteria~\cite{Vogel08,Miran10} and entanglement
criteria~\cite{Shchukin,Miran09}, based on moments of the
annihilation and creation operators, as discussed in Secs. II
and III, it is possible to construct infinitely many
nonclassicality and entanglement witnesses. These witnesses,
after truncation according to Eq.~(\ref{redefine}), can
exhibit the SV and SR effects when analyzing their time
evolution.

We hope that these observations might motivate deeper
analysis of SV and SR of various nonclassicality witnesses in
specific models and also in experimental scenarios.

\begin{acknowledgments}
We thank Prof.~Ryszard Tana\'s for discussions. A.M.
acknowledges support from the Polish Ministry of Science and
Higher Education under Grant No.~2619/B/H03/2010/38. X.W. is
supported by National Natural Science Foundation of China
(NSFC) with Grants No.~11025527, No.~10874151, and
No.~10935010. Y.X.L. was supported by the National Natural
Science Foundation of China under Grant No. 10975080. F.N.
acknowledges partial support from the Laboratory of Physical
Sciences, National Security Agency, Army Research Office,
DARPA, AFOSR, National Science Foundation Grant No.~0726909,
JSPS-RFBR Contract No.~09-02-92114, Grant-in-Aid for
Scientific Research (S), MEXT Kakenhi on Quantum Cybernetics,
and Funding Program for Innovative R\&D on S\&T (FIRST).
\end{acknowledgments}

\end{document}